\documentclass{amsart}
\input epsf

\newtheorem{theorem}{Theorem}[section]

\theoremstyle{definition}

\theoremstyle{remark}
\newtheorem{remark}[theorem]{Remark}

\usepackage{graphicx}
\usepackage{color}
\usepackage{amsmath}
\usepackage{amsfonts}
\usepackage{amssymb}
\newcommand{\be}{\begin{equation}}
\newcommand{\ee}{\end{equation}}

\newcommand{\ba}{\begin{array}}

\newcommand{\ea}{\end{array}}

\newcommand{\beq}{\begin{eqnarray}}

\newcommand{\eeq}{\end{eqnarray}}

\newtheorem{lm}{lemma}

\newtheorem{thee}{theorem}

\newtheorem{proo}{proposition}

\newtheorem{co}{corollary}

\newtheorem{rem}{remark}

\newtheorem{deff}{definition}

\newcommand{\bd}{\begin{deff}}

\newcommand{\ed}{\end{deff}}

\newcommand{\bl}{\begin{lm}}

\newcommand{\el}{\end{lm}}

\newcommand{\bp}{\begin{proo}}

\newcommand{\ep}{\end{proo}}

\newcommand{\bt}{\begin{thee}}

\newcommand{\et}{\end{thee}}

\newcommand{\bc}{\begin{co}}

\newcommand{\ec}{\end{co}}

\newcommand{\brm}{\begin{rem}}

\newcommand{\erm}{\end{rem}}

\usepackage{t1enc}

\newcommand{\newc}{\newcommand}

\let\ccdot\cdot
\def\cdot{\hbox to 2.5pt{\hss$\ccdot$\hss}}

\newc{\aR}{\mbox{\boldmath{$ R$}}}
\newc{\aS}{\mbox{\boldmath{$ S$}}}
\newc{\aT}{\mbox{\boldmath{$ T$}}}
\newc{\aW}{\mbox{\boldmath{$ W$}}}

\newc{\aK}{\mbox{\boldmath{$ K$}}}
\newc{\aL}{\mbox{\boldmath{$ L$}}}

\usepackage{amssymb}
\usepackage{amscd}

\newc{\obstrn}[2]{B^{#1}_{#2}}

\newcommand{\rpl}                         
{\mbox{$
\begin{picture}(12.7,8)(-.5,-1)
\put(0,0.2){$+$}
\put(4.2,2.8){\oval(8,8)[r]}
\end{picture}$}}

\newcommand{\lpl}                         
{\mbox{$
\begin{picture}(12.7,8)(-.5,-1)
\put(2,0.2){$+$}
\put(6.2,2.8){\oval(8,8)[l]}
\end{picture}$}}

\usepackage{ifthen}

\newcommand{\bbR}{\mathbb{R}}

\newc{\tensor}[1]{#1}
\newc{\Mvariable}[1]{\mbox{#1}}
\newc{\down}[1]{{}_{#1}}
\newc{\up}[1]{{}^{#1}}

\newc{\JulyStrut}{\rule{0mm}{6mm}}
\newc{\midtenPan}{\mbox{\sf S}}
\newc{\midten}{\mbox{\sf T}}
\newc{\midtenEi}{\mbox{\sf U}}
\newc{\ATen}{\mbox{\sf E}}
\newc{\BTen}{\mbox{\sf F}}
\newc{\CTen}{\mbox{\sf G}}

\def\sideremark#1{\ifvmode\leavevmode\fi\vadjust{\vbox to0pt{\vss
 \hbox to 0pt{\hskip\hsize\hskip1em
 \vbox{\hsize3cm\tiny\raggedright\pretolerance10000
 \noindent #1\hfill}\hss}\vbox to8pt{\vfil}\vss}}}%

\newcommand{\Span}{\mathrm{Span}}

\numberwithin{equation}{section}

\newcounter{romenumi}
\newcommand{\labelromenumi}{(\roman{romenumi})}

\begin{document}
\title{Split octonions and Maxwell equations}

\author{Pawe\l~ Nurowski} \address{Instytut Fizyki Teoretycznej,
Uniwersytet Warszawski, ul. Hoza 69, Warszawa, Poland}
\email{nurowski@fuw.edu.pl} 

\date{\today}

\begin{abstract} A formulation of the Maxwell equations in terms of the split 
octonions is presented.
\end{abstract}
\maketitle
A well known multiplication rule for the imaginary octonions \cite{baez}, depicted schematically on the Fano plane,\\
\centerline{\epsfxsize=5cm \epsfbox{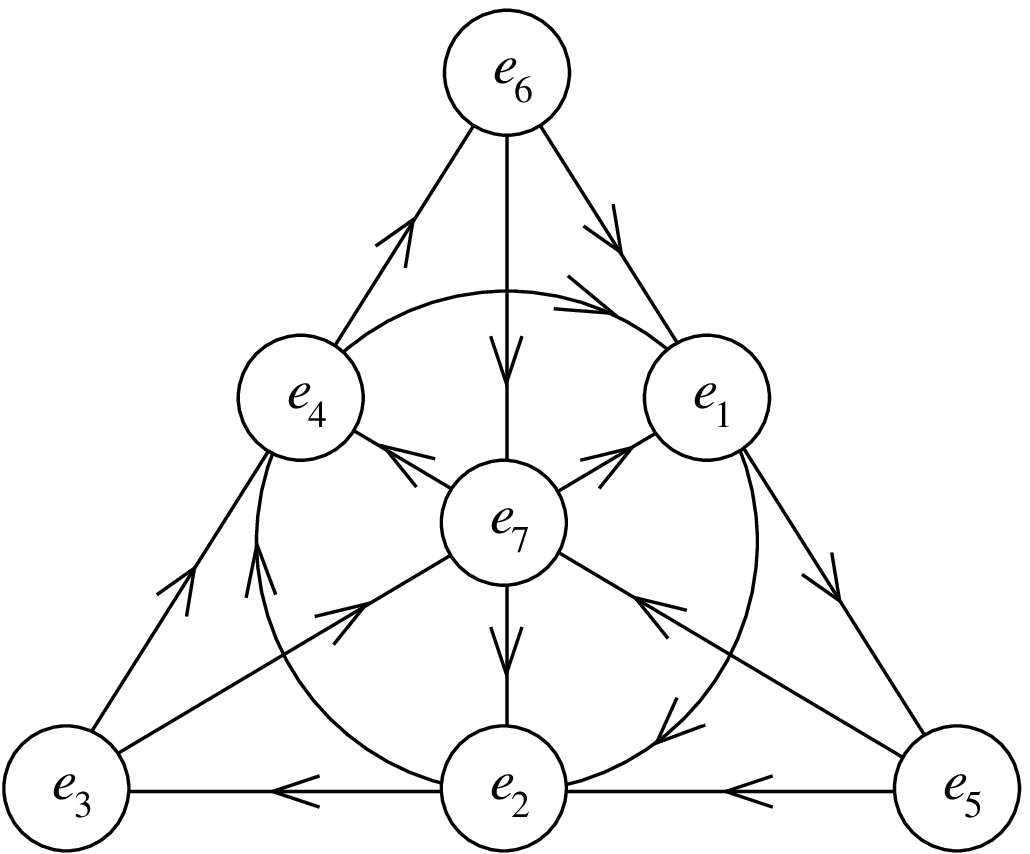}}
may be generalized to the following multiplication table    
\begin{center}   
{\small   
\begin{tabular}{|c|c|c|c|c|c|c|c|c|}                    \hline   
      & $e_1$ & $e_2$ & $e_3$ & $e_4$  & $e_5$ & $e_6$ & $e_7$ \\ \hline   
  $e_1$ & $-1$  & $e_4$ & $e_7$ & $-e_2$ & $e_6$ & $-e_5$ & $-e_3$ \\ \hline   
$e_2$ & $-e_4$ & $-1$ & $e_5$ & $e_1$ & $-e_3$ & $e_7$ & $-e_6$     \\ \hline   
$e_3$ & $-e_7$ & $-e_5$ & $\mp 1$ & $e_6$ & $\pm e_2$ & $\mp e_4$ & $\pm e_1$   \\ \hline   
$e_4$ & $e_2$ & $-e_1$ & $-e_6$ & $-1$ & $e_7$ & $e_3$ & $-e_5$   \\ \hline   
$e_5$ & $-e_6$ & $e_3$ & $\mp e_2$ & $-e_7$ & $\mp 1$ & $\pm e_1$ & $\pm e_4$    \\ \hline   
$e_6$ & $e_5$ & $-e_7$ & $\pm e_4$ & $-e_3$ & $\mp e_1$ & $\mp 1$ & $\pm e_2$     \\ \hline   
$e_7$ & $e_3$ & $e_6$ & $\mp e_1$ & $e_5$ & $\mp e_4$ & $\mp e_2$ & $\mp 1$    \\  \hline   
\end{tabular}}. 
\end{center} 
In this table the upper sign corresponds to the multiplication of the imaginary octonions 
and the lower sign to the multiplication of the imaginary split octonions. Via 
the above multiplication rules the 8-dimensional real vector space 
$$\Span_\bbR\{1,e_1,e_2,e_3,e_4,e_5,e_6,e_7\}$$
gets equipped with the structure of two 
(nonassociative) algebras with unit $1$. These two algebras are called 
octonions for the upper sign in the table and the split octonions for the lower sign.

In this letter we will use a vector-valued (split)octonion $\vec{Q}=(e_1,e_2,e_4)$ and its corresponding vector-valued (split)octonion $e_7\vec{Q}=(e_7e_1,e_7e_2,e_7e_4)=(e_3,e_6,e_5)$.

Consider now two vector fields on $\bbR^4$
$$\vec{E}:\bbR^4\to\bbR^3,\quad\quad (t,x,y,z)\mapsto\vec{E}=(E_x,E_y,E_z)$$ and 
$$\vec{B}:\bbR^4\to\bbR^3,\quad\quad(t,x,y,z)\mapsto\vec{B}=(B_x,B_y,B_z).$$ Define
\begin{eqnarray*}F&=&E_x e_1+E_y e_2+E_z e_4+ B_x e_3 + B_y e_6+B_z e_5=\vec{E}\vec{Q}+\vec{B}(e_7\vec{Q})\\
&=&(E_x+e_7 B_x)e_1+(E_y+e_7 B_y)e_2+(E_z+e_7 B_z)e_4
=(\vec{E}+e_7\vec{B})\vec{Q}.
\end{eqnarray*}
and 
$$\partial=e_1\partial_x+e_2\partial_y+e_4\partial_z+e_7\partial_t.$$

Then using the above multiplication table and the standard notation of vector calculus in $\bbR^3$ we have:
\begin{eqnarray*}\partial F=-\vec{\nabla}\vec{E}
+(\vec{\nabla}\times\vec{E}\mp\tfrac{\partial\vec{B}}{\partial t})\vec{Q}
+(-\vec{\nabla}\times\vec{B}+\tfrac{\partial\vec{E}}{\partial t})(e_7\vec{Q})+(\vec{\nabla}\vec{B})e_7.
\end{eqnarray*}
Thus we see that if we choose the split octonions (lower sign) then
$$\partial F=0$$
is equivalent to the equations
\begin{eqnarray*}\vec{\nabla}\vec{E}&=&0,\quad\quad
\vec{\nabla}\times\vec{E}=-\tfrac{\partial\vec{B}}{\partial t}\\
\vec{\nabla}\vec{B}&=&0,\quad\quad\vec{\nabla}\times\vec{B}=\tfrac{\partial\vec{E}}{\partial t}.
\end{eqnarray*}
Several remarks are in order:
\begin{remark}
Why the Nature prefers the split octonions rather than the octonions
for the electromagnetism? 
\end{remark}
\begin{remark}
It is interesting to note that $F$ defined above is not a generic imaginary split octonion. If we were to chose a generic imaginary split octonion
$$F= E_x e_1+E_y e_2+E_z e_4+ B_x e_3 + B_y e_6+B_z e_5+S e_7,$$
where $S=S(t,x,y,z)$ was arbitrary function on $\bbR^4$, then 
\begin{eqnarray*}\partial F=(-\vec{\nabla}\vec{E}+\tfrac{\partial S}{\partial t})
+(\vec{\nabla}\times\vec{E}+\tfrac{\partial\vec{B}}{\partial t})\vec{Q}
+(-\vec{\nabla}\times\vec{B}+\tfrac{\partial\vec{E}}{\partial t}-\vec{\nabla}S)(e_7\vec{Q})+(\vec{\nabla}\vec{B})e_7.
\end{eqnarray*}
In such case $\partial F=0$ would correspond to
\begin{eqnarray*}\vec{\nabla}\vec{E}&=&\tfrac{\partial S}{\partial t},\quad\quad
\vec{\nabla}\times\vec{E}+\tfrac{\partial\vec{B}}{\partial t}=0\\
\vec{\nabla}\vec{B}&=&0,\quad\quad-\vec{\nabla}\times\vec{B}+\tfrac{\partial\vec{E}}{\partial t}=\vec{\nabla} S.
\end{eqnarray*}
These equations for $(\vec{E},\vec{B})$ could be then interpreted as the Maxwell equations for electromagnetic field $(\vec{E},\vec{B})$ generated by the charge density $\rho=\tfrac{\partial S}{\partial t}$ and the current density $\vec{j}=\vec{\nabla}S$. Note that to get the magnetic charge densities and magnetic currents we would need to introduce the generic (not purely imaginary) split octonion $F$.
\end{remark}
\begin{remark}
Now the story is quite puzzling: an authomorphism $\sigma$ of the split octonions is an element of the noncompact real form of the exceptional Lie group $G_2$.  Since $\sigma(\partial F)=\sigma(\partial)\sigma(F)$ then, if $F$ satisfied the Maxwell equations $\partial F=0$,  the transformed $\sigma(F)$ would satisfy the equations $\sigma(\partial)\sigma(F)=0$. But the transformed field $\sigma(F)$ is a general split  octonion; the transformed derivative $\sigma(\partial)$ also is. The physical interpretation of the $\sigma$-induced transformation on the space time coordinates $(t,x,y,z)$ and the electromagnetic field  $(\vec{E},\vec{B})$ would be interesting.  
\end{remark}
 
\end{document}